\documentclass[aps,prb,preprint]{revtex4}

\usepackage{graphicx}
\usepackage{graphics}
\usepackage{epsfig}

\begin{document}
\title{ Suppression of superfluid density in the superfluid-supersolid transition }

\author{Tai-Kai Ng}
\address{Department of Physics, Hong Kong University of Science and
Technology, Clear Water Bay Road, Hong Kong}

\begin{abstract}

  We show that the rather unexpected pressure dependence of superfluid density observed near the superfluid-supersolid
 transition by Kim {\em et.al.}[M.H.W. Chan, {\em private communication}], can be understood if the transition from
 superfluid to supersolid state is a second order or weakly first order transition from the superfluid state to a
 super-CDW state with non-uniform Bose-condensation amplitude. The suppression of superfluid density is a direct
 consequence of softening of phonon mode at finite wave-vector $|\vec{Q}|\sim Q_0$ around the quantum phase
 transition.

  \end{abstract}

\pacs{75.10.Jm, 05.30.Jp, 74.25.Dw}

\maketitle

\narrowtext

   Following the basic theoretical understanding of superfluidity in in $He^4$, the possibility of a so-called
 supersolid state has been proposed\cite{t1,t2,t3,t4}. In its simplest form, the theoretically proposed
 supersolid state can be thought of as a state with spatially non-uniform Bose-condensation amplitude
 $\bar{\psi}(\vec{x})$\cite{t1,t2,t3,t4}. For example,
 \begin{equation}
 \label{con1}
 \bar{\psi}(\vec{x})=b_0+b_Q\cos(\vec{Q}.\vec{x})
 \end{equation}
 $(b_0>b_Q)$, corresponding to a Bose-condensed charge-density wave, or super-CDW, state. Notice that this state,
 if stable, has a non-zero superfluid density by construction, and is very different from the usual insulator state
 where superfluid density vanishes.

  The supersolid state is observed recently in a torsional oscillator experiment by Kim
  {\em et.al.}\cite{chan1, chan2} where finite (but small) superfluid density was observed in an otherwise solid
  $He^4$ state. A rather unexpected finding of the experiment is that the superfluid density is not a monotonic
  function of pressure but increases first as pressure increases, and decreases again when pressure is higher,
  whereas naively we expect that the superfluid density should be a monotonic decreasing function with
 pressure\cite{moses}. A number of theoretical works were triggered by the discovery of this new
 state\cite{t5,t6,t7,t8,t9,t10}. The purpose of our paper is to point out that the rather unexpected pressure
 dependence of superfluid density as found in the experiment can be understood if we identify the transition from
 superfluid to supersolid state as a second order or weakly first order phase transition from the superfluid state
 to a super-CDW state, whereas the transition to the real insulator state occurs at a higher pressure. The
 suppression of the superfluid density $\rho_s$ is a direct consequence of enhanced thermal fluctuation coming
 from softening of phonon mode at finite wave-vector $|\vec{Q}|\sim Q_0$ around the quantum phase
 transition and is quite independent of the microscopic details of the system.

  To demonstrate we consider an effective action for the boson system of form
 \begin{equation}
 \label{ham}
  S\sim \int_0^{\beta}d\tau\left(\int{d}^dx\left[i\rho_1\partial_{\tau}\theta+
  {\bar{\rho}\hbar^2\over2m}(\nabla\theta)^2\right]+V[\rho]\right),
 \end{equation}
 where $\rho=\bar{\rho}+\rho_1$ and $\theta$ are the density and phase variables characterizing the boson system and
 $V[\rho]$ is an energy functional. The ground state density profile $\bar{\rho}$ of the system is determined by
 minimizing $V[\rho]$ with respect to $\rho$. $\rho_1$ describes the density fluctuations above the ground state.
 The boson creation (annihilation) operators are related to the density-phase variables by
 $\psi^+(\psi)(\vec{x})=\sqrt{\rho(\vec{x})}e^{-(+)i\theta(\vec{x})}$. Specifically,
 \begin{equation}
 \label{gauss}
 V[\rho]=\int{d}^3x{\hbar^2\over2m}(\nabla\sqrt{\rho})^2+U[\rho],
 \end{equation}
 in usual interacting boson problem where
 $U[\rho]={1\over2}\int d^3x\int d^3x'\rho(\vec{x})U(|\vec{x}-\vec{x}'|)\rho(\vec{x}')$ is the potential energy of
 the bosons with $U(x)$ being the corresponding interaction potential. Here we shall treat $V[\rho]$ as a
 phenomenological effective potential with which the ground state and low energy properties of the system
 can be described by the Gaussian theory.

   In the normal superfluid state, the density is uniform and $\bar{\rho}(\vec{x})=n$. Small density fluctuations
 around the ground state can be described by expanding $V[\rho=n+\Delta\rho]$ in a power series of
 $\Delta\rho$\cite{slt}, i.e.
 \begin{eqnarray}
 \label{veff}
 V[\rho] & \sim & V[n]+\sum_{\vec{Q}}A(|\vec{Q}|)\Delta\rho_{\vec{Q}}\Delta\rho_{-\vec{Q}}+\sum_{\vec{Q}_1,\vec{Q}_2}
 B(\vec{Q}_1,\vec{Q}_2)\Delta\rho_{\vec{Q}_1}\Delta\rho_{\vec{Q}_2}\Delta\rho_{-\vec{Q}_1-\vec{Q}_2}
 \\  \nonumber
 & & +\sum_{\vec{Q}_1,\vec{Q}_2,\vec{Q}_3}C(\vec{Q}_1,\vec{Q}_2,\vec{Q}_3)\Delta\rho_{\vec{Q}_1}
 \Delta\rho_{\vec{Q}_2}\Delta\rho_{\vec{Q}_3}\Delta\rho_{-\vec{Q}_1-\vec{Q}_2-\vec{Q}_3}+....
 \end{eqnarray}
 where we assume that the system is translational invariant. Notice that the coefficients $A,B,C,...$ are in general
 functions of external parameters like pressure and temperature.

   For $B=0$, the transition from the normal superfluid state to super-CDW state is characterized by a change of sign
 of $A(Q)$ from positive to negative at some wavevector $Q=Q_0>0$. The change in $A(Q)$ can be induced
 by pressure as in the experiment of Chan {\em et.al.}\cite{chan1,chan2} or through other means. In this case the
 quantum phase transition is a second order phase transition. For $B\neq0$, the transition becomes a first order
 phase transition occurring at finite value of $A(Q_0)>0$. In the following, we shall show that complete
 suppression of superfluid density at the quantum phase transition is a natural consequence of the above
 phenomenological Lagrangian when $B=0$, and suppression of superfluid density still occurs when $B$ is small,
 when the transition is weakly first order. Notice that for the super-CDW state to be stable, $A(Q)$ must has the
 property that it is less than zero only at a small region around $Q\sim Q_0$, and is positive otherwise.

  To get some physical feeling of $A(Q)$, we consider $V[\rho]$ of form \ (\ref{gauss}). In this case, it is easy to
 show that $A(Q)=(\epsilon_{\vec{Q}}+2nU(Q))/4n$, where $\epsilon_{\vec{q}}=\hbar^2q^2/2m$ is the kinetic energy and
 $U(q)=\int{d}^dxe^{i\vec{q}.\vec{x}}U(x)$ is the Fourier transform of the inter-particle interaction $U(x)$.
 In terms of the inter-particle potential $U(q)$, the stability requirement of the CDW state means that we must have
 $U(q\rightarrow0)>0$ with an absolute minimum at $q=Q_0\sim$ {\em (particle spacing)}$^{-1}$, with
 $\epsilon_{\vec{Q}_0}+2nU(Q_0)<0$ around this region.

    It is interesting to note that this form of effective interaction potential has been used to describe low
 temperature properties of helium liquid\cite{pines} and can be derived from, for example, the STLS scheme\cite{ng1}.
 In these theories, $U(Q_0)<0$ leads to a peak in the structure factor $S(q\sim Q_0)$ and a corresponding dip in the
 phonon spectrum $E(q\sim Q_0)$. As pressure increases, $U(Q_0)$ decreases further and the peak magnitude in
 $S(Q_0)$ increases. The increase in magnitude of $S(Q_0)$ is interpreted as a precursor to formation of Wigner
 crystal\cite{pines,ng1}. Notice that correspondingly the phonon energy $E(q\sim Q_0)$ softens. The
 superfluid-super-CDW instability is driven by an instability in the phonon spectrum at $q\sim Q_0$ when
 $E(Q_0)\rightarrow0$ and $S(Q_0)$ diverges.

   We next analyze the phenomenological Lagrangian \ (\ref{ham}) and \ (\ref{veff}) in Gaussian approximation. First
 we consider $B=0$ and $A(Q)>0$. In this case higher order terms in \ (\ref{veff}) are unimportant. After
 integrating out the $\rho_1$ (or $\Delta\rho)$ fields, we obtain
 \begin{equation}
 \label{a0}
  S_{\theta}\sim\int_0^{\beta}d\tau\int{d}^dq\left(n\epsilon_{\vec{q}}(\theta(\vec{q}))^2+
 {1\over A(q)}(\partial_{\tau}\theta(\vec{q}))^2\right).
 \end{equation}
  where the usual phonon dispersion $E(\vec{q})=\sqrt{n\epsilon_{\vec{q}}A(q)}\sim
 \sqrt{\epsilon_{\vec{q}}\left(\epsilon_{\vec{q}}+2nU(\vec{q})\right)}$ follows. As $\vec{q}\rightarrow0$,
 $E(\vec{q})\rightarrow\sqrt{nA(0)\epsilon_{\vec{q}}}\sim|\vec{q}|$, which is the usual superfluid Goldstone mode.
 We also observe that $E(Q_0\rightarrow0)\rightarrow0$ when $A(Q_0)\rightarrow0$, corresponding to softening of
 the phonon spectrum at the $|\vec{Q}|=Q_0$ when the CDW instability occurs.

  It is straightforward to compute the depletion in superfluid density near the superfluid-super-CDW transition
  coming from thermal fluctuations. The superfluid density is given by $\rho_s(T)=n-\rho_n(T)$, where
  \begin{equation}
  \label{depletion}
    \rho_n(T)=-{1\over3}\int{d^3q\over(2\pi)^3}(\vec{q})^2{dn_B(\epsilon)\over d\epsilon}\mid_{\epsilon=E(\vec{q})},
  \end{equation}
  is the normal liquid density\cite{kha}. $n_B(E)=(e^{E\over k_BT}-1)^{-1}$ is the boson occupation number.

   To proceed further we assume $E(q)^2\sim E(Q_0)^2+c^2(q-Q_0)^2$ around $q\sim Q_0$, where
 $c^2=d^2E(q)/dq^2|_{q=Q_0}>0$ and $E(Q_0)(A(Q_0))\rightarrow0$. The contribution to $\rho_n$ from around
 $|\vec{Q}|\sim Q_0$ can be estimated by evaluating the integral. For $k_BT>>E(Q_0)$ we obtain
 \begin{equation}
 \label{depletion1}
 \rho_n(T)\sim{Q_0^4\over6\pi^2c}{k_BT\over E(Q_0)}\left(\tan^{-1}{k_BT\over E(Q_0)}\right)+\rho_n^{ph}(T),
 \end{equation}
 where $\rho_n^{ph}(T)\sim T^4$ is the contribution from ordinary phonon around zero momentum and
 \begin{equation}
 \label{depletion2}
 \rho_n(T)\sim {Q_0^4\over6\pi c}\sqrt{2E(Q_0)\over\pi k_BT}e^{-E(Q_0)\over k_BT}+\rho_n^{ph}(T),
 \end{equation}
 for $k_BT<<E(Q_0)$. We observe that $\rho_n(T)$ diverges as $k_BT/E(Q_0)$ at the superfluid-super-CDW transition
 when $k_BT>>E(Q_0)\rightarrow0$, indicating that superfluidity is destroyed as a result of thermal fluctuations at
 the critical point at any non-zero temperature $T\neq0$. The suppression of superfluidity is analogous to the
 suppression of Bose-condensation by thermal fluctuation in one-dimensional Bose gas. Notice that the suppression
 occurs because the instability occurs at finite wave-vector $Q_0$ and would be absence in transitions where the
 instability occurs at $\vec{Q}=0$ or in lattice models where $\vec{Q}$ commensurate with the underlying lattice
 structure.

  The $k_BT/E(Q_0)$ singularity is cutoff by a first-order phase transition when $B\neq0$. To see how this
  occurs we consider the potential energy \ (\ref{veff}), assuming that $A(Q)$ approaches zero at $Q=Q_0$ and the
  boson system forms a super-CDW state with period $\sim Q_0^{-1}$. At the vicinity of the transition the CDW wave
  can be viewed as superposition of density waves with same amplitude but travelling at different directions with
  different phases\cite{slt}, i.e., $\rho(\vec{x})\sim\sum_{\vec{G}}\Delta\rho_{\vec{G}}e^{i\vec{G}.\vec{x}}$
  where $\vec{G}$'s are primary reciprocal lattice vectors with $|\vec{G}|=Q_0$ and $\Delta\rho_{\vec{G}}
  \sim|\Delta\rho_0|e^{i\theta_{\vec{G}}}$. In this case, the potential energy \ (\ref{veff}) becomes
  \begin{equation}
  \label{vc}
  V[\rho]\rightarrow V[n]+MA(Q_0)(\Delta\rho_0)^2+\bar{B}(Q_0)(\Delta\rho_0)^3+\bar{C}(Q_0)(\Delta\rho_0)^4+....
  \end{equation}
  where $M=$ number of reciprocal lattice vectors $\vec{G}$'s that enters the construction of the CDW
  state, $\bar{B}\sim\sum_{\vec{G}'s}B(\vec{G}_1,\vec{G}_2)e^{i(\theta_1+\theta_2)}$ and
  $\bar{C}\sim\sum_{\vec{G}'s}C(\vec{G}_1,\vec{G}_2,\vec{G}_3)e^{i(\theta_1+\theta_2+\theta_3)}>0$ are constants that
  depends on $Q_0, M$ and the lattice structure of the CDW state.

    Minimizing $V[\rho]$ up to fourth order terms, it is easy to see that $\Delta\rho_0\sim\bar{B}/2\bar{C}>0$ when
  $MA\leq{\bar{B}^2\over4\bar{C}}$, and the second order phase transition is cutoff by a first order phase transition
  when $\bar{B}\neq0$. Correspondingly, the $k_BT/E(Q_0)$ divergence of $\rho_n(T)$ computed in
  Eq.\ (\ref{depletion1}) is cutoff with
    \[
      E(Q_0)\rightarrow\sqrt{{n\epsilon_{\vec{Q}_0}\bar{B}^2\over4M\bar{C}}}.  \]

    Notice that the superfluid density changes discontinuously across the phase transition as a result of the
  discontinuous change in the excitation spectrum when $\bar{B}\neq0$. The superfluid density in the super-CDW state
  can be written as $\rho_s^{CDW}(T)=\rho_s^{CDW}(0)-\rho_n^{CDW}(T)$, where $\rho_s^{CDW}(0)\sim n-
  \alpha(\Delta\rho_0)^2$ and $\rho_n^{CDW}(T)\sim\rho_n(T)+\beta(T)(\Delta\rho_0)^2$ around the phase transition,
  where $\alpha,\beta(T)$ are non-diverging constants depending on the CDW lattice structure. It
  is easy to show that at the CDW side of the transition, $V[\rho]\sim V[n]+V[\Delta\rho_0]+A'\rho_1^2$, where
  $A'=\bar{B}^2/4\bar{C}$. Therefore we expect that the low energy cutoff is replaced by
   \[
      E_{CDW}(Q_0)\sim\sqrt{{n\epsilon_{\vec{Q}_0}\bar{B}^2\over4\bar{C}}}  \]
  in this case. Phonon excitations with energy above $E_{CDW}(Q_0)$ is only weakly affected by the presence of
  $\Delta\rho_0\neq0$ and $\rho_n(T)$ is given by Eq.\ ({\ref{depletion1}) and (\ref{depletion2}), except that
  $E(Q_0)$ is replaced by $E_{CDW}(Q_0)$.

    When applying our results to the superfluid-supersolid transition, it suggests that what
   Kim {\em et.al.}\cite{moses} have observed experimentally may be the depletion of superfluid density
   around the quantum phase transition due to thermal fluctuations. Moreover, it suggests that similar depletion
   will be found in the {\em superfluid} side of the transition. The depletion is temperature dependent and there
   will be a discontinuity in the superfluid density across the transition if the transition is first order, with the
   superfluid density in the {\em normal superfluid} side being suppressed more, at least when $\bar{B}$ is small
   enough. The phonon dispersion at $Q\sim Q_0$ hardens\cite{t5} and the superfluid density rises again when the
   system moves away from both sides of the critical point. Notice equations (7) and (8) predict also a rather
   specific temperature dependence of superfluid density around the transition. Therefore, a measurement of
   temperature-dependent superfluid density across the transition can determine the order of the transition, and
   provides a test to our theory.

 {\it Acknowledgements}
  The author thank Dr. Yi Zhou for reading the manuscript and for his helpful comments. This work is
  supported by HKRGC through Grant 602803.


\references
 \bibitem{t1} A.F. Andreev and I.M. Lifshitz, {\em Sov. Phys. JETP}{\bf 29}, 1107 (1969).
 \bibitem{t2} G.V. Chester, \pra {\bf 2}, 256 (1970).
 \bibitem{t3} A.J. Leggett, \prl {\bf 25}, 1543 (1970).
 \bibitem{t4} H. Matsuda and T. Tsuneto, {\em Prog. Theor. Phys.} {\bf 46}, 411 (1970).
 \bibitem{t5} D.L. Kovrizhin, G.V. Pai and S. Sinha, cond-mat/0410512.
 \bibitem{t6} X. Dai, M. Ma and F.C. Zhang, cond-mat/0501373.
 \bibitem{t7} N. Kumar, cond-mat/0507553.
 \bibitem{t8} G. Baskaran, cond-mat/0505160.
 \bibitem{t9} M. Boninsegni and N. Prokof'ev, cond-mat/0507620.
 \bibitem{t10} A.T. Dorsey, P.M. Goldbart and J. Toner, cond-mat/0508271.
 \bibitem{chan1} E. Kim and M.H.W. Chan, {\em Nature} {\bf 427}, 225(2004).
 \bibitem{chan2} E. Kim and M.H.W. Chan, {\em Science} {\bf 305}, 1941(2004).
 \bibitem{moses} M.H.W. Chan, private communication;(see also http://www.wycamp.ust.hk/).
 \bibitem{slt} see for example, P. Bak, \prl {\bf 54}, 1517 (1985).
 \bibitem{pines} C.H. Aldrich and D. Pines, {\em J. Low Temp. Phys.} {\bf 32}, 689 (1978).
 \bibitem{ng1} T.K. Ng and K.S. Singwi, \prb {\bf 35}, 1708 (1987).
 \bibitem{kha} see for example, I.M. Khalatinikov, {\em An Introduction to the Theory of Superfluidity},
   Addison-Wesley 1989.
 \end{document}